\documentstyle[aps,prbbib,twocolumn,epsf]{revtex}

\begin{document}
\draft
\title{Current partition: Nonequilibrium Green's function Approach}

\author{Baigeng Wang$^1$, Jian Wang $^1$ and Hong Guo$^{1,2}$}
\address{1. Department of Physics, The University of Hong Kong, 
Pokfulam Road, Hong Kong, China\\
2. Center for the Physics of Materials and Department 
of Physics, McGill University, Montreal, PQ, Canada H3A 2T8\\}
\maketitle

\begin{abstract}
{\bf 
We present a solution to the problem of AC current partition in a 
multi-probe mesoscopic conductor within the nonequilibrium Green's function 
formalism. This allows the derivation of dynamic conductance 
which is appropriate for nonequilibrium situations and which satisfies 
the current conservation and gauge invariance requirements. This formalism 
presents a significant generalization to previous theory:
(i) there is no limit in the frequency, and (ii) it allows 
detailed treatments of interactions in the mesoscopic region. 
The formalism is applied to calculate dynamic conductance of tunneling 
structures with and without assuming wideband limit.
}
\end{abstract}

\pacs{73.23.Ad,73.40.Gk,72.10.Bg}

The problem of electric current partition in a mesoscopic multi-probe
conductor is a fundamental issue of quantum transport. In the
familiar DC situation
the issue is well understood from the
Landauer-B\"uttiker formulation\cite{landauer}. Under time varying 
AC conditions, this problem becomes more complicated due to the 
presence of displacement current which is induced by the AC fields
according to electrodynamics\cite{but0}. In this case,
one needs to know how to partition the displacement current in 
addition to the particle current for each probes of the conductor.
Without this knowledge one cannot obtain correct results of dynamic
conductance: the electric current will not be conserved unless
the displacement current is taken into account. 
There is another fundamental requirement in the transport which is the
gauge invariance of the theory. This simply means that the physics 
depends only on the voltage difference thus shifting voltages everywhere 
by the same constant amount should not not alter the results. In the
AC transport the gauge invariance will not be satisfied without taking
into account of displacement current. 

Quantum transport 
under dynamic conditions is the focus of a number of recent 
experiments\cite{price,webb,vegvar,chen,oosterkamp}. 
Under low frequency AC fields the system is near equilibrium, the 
problem of current partition can be analyzed using scattering 
matrix theory (SMT). B\"uttiker and co-workers have developed a 
theoretical formalism\cite{but1,but2} based on SMT to analyze linear AC 
transport and derived the dynamic conductance $G_{\alpha\beta}(\omega)$. 
Here a subscripts denotes a probe. By including the contribution of 
displacement current and its partition among the probes, the derived 
dynamic conductance guarantees current conservation and gauge 
invariance\cite{but1}. Obviously, displacement current should be 
even more important for higher frequencies and for situations far 
from equilibrium. In this Letter we address the 
problem of current partition {\it far from equilibrium}, and with 
this information we derive a dynamic conductance expression 
which conserves electric current and satisfies the gauge invariance
under nonequilibrium conditions.

Far from equilibrium, the Keldysh nonequilibrium Green's 
function\cite{keldysh,ng,wingreen,pastawski,datta,chen1,hettler,mahan}
(NEGF) has been widely applied and many problems of great interests have been 
analyzed. However, within this formalism contribution of displacement current 
has not been included. In other words the current 
conservation and gauge invariant condition have not been 
satisfied\cite{stafford}. A direct consequence of neglecting displacement 
current is to predict incomplete results for dynamic conductance, such as 
that for a parallel plate capacitor. The fact of violating current 
conservation within the nonequilibrium theories has been 
recognized in a number of publications\cite{datta,hettler,jauho3}. The problem 
is not solved because of a lack of knowledge for partition the 
displacement current.

Let's consider a quantum coherent multi-probe conductor with the Hamiltonian 
\begin{eqnarray}
H &=& \sum_{k\alpha} \epsilon_{k\alpha}(t) c^{\dagger}_{k\alpha}
c_{k\alpha} + H_{cen}\{d_n,d^{\dagger}_n\} \nonumber \\
&+& \sum_{k\alpha, n} [T_{k\alpha,n} c^{\dagger}_{k\alpha} d_n + c.c.]
\label{Ham}
\end{eqnarray}
where $\epsilon_{k\alpha}(t) = \epsilon^0_{k\alpha} + qV_{\alpha} 
\cos{\omega t} $. The first term is the Hamiltonian of probes where the 
AC signal is applied; 
the second term is the general 
Hamiltonian for the scattering region which is a polynomial in 
$\{d^{\dagger}_n,d_n \}$ that commutes with the electron number
operator\cite{stafford} $N=\sum_n d^{\dagger}_n d_n$;
the coupling between
the probes and the scattering region is given by the last term 
with coupling matrix $T_{k\alpha,n}$. 
$c^{\dagger}_{k\alpha}$ ($c_{k\alpha}$) is the creation (annihilation) 
operator inside the $\alpha$-probe, $d^{\dagger}_n$ ($d_n$) 
is for the scattering region. 

The {\it particle current} (conduction current) inside 
probe $\alpha$ can be calculated 
using the equation of motion for the number of carriers inside that 
probe. In terms of the Green's function and the self-energy, it 
can be written in the familiar form\cite{wingreen,datta} in 
frequency representation ($\hbar =1$),
\begin{eqnarray}
& &I^c_{\alpha}(\omega) = -q \int^{+\infty}_{-\infty} \frac{dEdE_1}{(2\pi)^2} 
Tr\left[G^r(E+ \omega,E_1) ~\Sigma^<_{\alpha}(E_1,E)\right.\nonumber \\
& & +G^<(E+ \omega,E_1) ~ \Sigma^a_{\alpha}(E_1,E)
~-~\Sigma^<_{\alpha}(E+ \omega,E_1) ~ G^a(E_1,E) \nonumber \\
& & \left. -\Sigma^r_{\alpha}(E+ \omega,E_1) ~ G^<(E_1,E) \ \right]
\label{i1}
\end{eqnarray}
where the Green's function and self-energy are defined in the usual 
manner\cite{wingreen,datta}. The double-time Fourier transform is defined as
$$F(E_1,E_2)=\int dt_1 \int dt_2 F(t_1,t_2) \exp[i(E_1 t_1 - E_2 t_2)]$$
We will be interested in the current component linear in voltage\cite{foot}, 
hence we simplify Eq.(\ref{i1}) by keeping the first order in voltage. 
The dynamic conductance coefficient $G_{\alpha\beta}^c(\omega)$ due to the 
particle current is defined as
$I^c_{\alpha}(\omega) = \sum_{\beta} G^c_{\alpha \beta}(\omega) 
V_{\beta}(\omega)$.
$G^c_{\alpha \beta}(\omega)$ is just the frequency dependent 
admittance matrix due to particle current. Without introducing
confusion, throughout the following we shall use notation 
$G$ with subscripts to denote conductance, while that without 
subscripts (unless otherwise stated) to denote Green's function.

In the presence of AC fields, the particle current alone is not 
conserved due to charge accumulation in the scattering region
$Q(t)$, {\it i.e.} $\sum_\alpha I_\alpha^c \neq 0$. It is the 
total current which is conserved\cite{datta},
$\sum_{\alpha} I^c_{\alpha}(t) + dQ(t)/dt = 0$ where
$I^d=dQ/dt$ is the displacement current. In Fourier space, 
\begin{equation}
\sum_{\alpha} I^c_{\alpha}(\omega) = i\omega Q(\omega)
\label{conserve}
\end{equation}
where
\begin{equation}
Q(\omega) = -\sum_{\beta} iq\int \frac{dE}{2\pi} ~ 
Tr[g^<_{\beta}(E+\omega,E)] ~ 
V_{\beta}
\label{charge}
\end{equation}
is the AC charge accumulation in the scattering region and
$g_\beta^<$ is the small-signal component of the Green's 
function $G^<$ defined as $\sum_\beta g^<_\beta V_\beta
=G^<-G^<_{eq}$ in the linear regime\cite{datta}, here $G^<_{eq}$ is for
equilibrium. This charge is related to the displacement field ${\bf D}$ of 
the internal Coulomb potential via Poisson equation\cite{but2} 
(more generally the potential satisfies Helmhotz equation): 
$\nabla \cdot {\bf D({\bf r},\omega)} = 4\pi \rho({\bf r},\omega)$ 
with $Q=\int \rho d{\bf r}$. Hence the displacement current
$-i(\omega/4\pi) \int {\bf D} d{\bf r}$ is just the time derivative of
the pile-up charge: $\sum_\alpha I_\alpha^d=dQ/dt$.
Thus the total current in probe $\alpha$ is 
$I_\alpha^{tot}=I_\alpha^c + I_\alpha^d$.  Current
conservation means $\sum_\alpha I_\alpha^{tot}=0$.

From Eq.(\ref{conserve}), only the total displacement current 
$I^d(\omega) = -i\omega Q(\omega)$ is known. We thus need to 
partition it into each probe $\alpha$. To give an example, the 
easiest situation would be the wideband limit 
where the coupling constants $\Gamma_\alpha$ are independent of energy. 
Hence, for an amount of
charge $Q$ in the scattering region, the probability of leaking out 
through probe $\alpha$ is simply $\Gamma_{\alpha}/\Gamma$ with
$\Gamma$ the total coupling between the probes to the scattering
region. This gives the displacement current partition for each probe:
$I^d_{\alpha} = (\Gamma_{\alpha}/\Gamma) I^d$.

More generally, we shall use two fundamental requirements of the theory 
to partition the displacement current: current conservation and gauge 
invariance.  Since current in probe
$\alpha$ is $I_{\alpha} = \sum_{\beta} G_{\alpha \beta} V_{\beta}$, 
current conservation means $\sum_\alpha G_{\alpha\beta}=0$; while gauge
invariance means $\sum_\beta G_{\alpha\beta}=0$. We partition the 
total displacement current $I^d$ into the contributions 
from individual probes $\alpha$ in the following form,
$I_{\alpha} = I^c_{\alpha} + A_{\alpha} I^d$, or 
$G_{\alpha \beta} = G^c_{\alpha \beta} + A_{\alpha} G^d_{\beta}$
where
\begin{equation}
G^d_{\beta} = -q\omega \int \frac{dE}{2\pi} 
~ Tr[g^<_{\beta}(E+\omega,E)]
\label{gd}
\end{equation}
is obtained from Eq. (\ref{charge}). The partition coefficients 
$A_\alpha$ must satisfy $\sum_{\alpha} A_{\alpha} =1$ to conserve the 
total current. To determine $A_\alpha$, we apply the gauge invariance 
condition and obtain 
$A_\alpha= -(\sum_{\gamma} 
G_{\alpha \gamma}^c)/(\sum_{\gamma}G^d_{\gamma})$. 
This gives the following form for the dynamic conductance
\begin{equation}
G_{\alpha \beta} = G^c_{\alpha \beta} - G^d_{\beta}
\frac{\sum_{\gamma} G^c_{\alpha \gamma}}{\sum_{\gamma} G^d_{\gamma}} 
\ \ .
\label{gab}
\end{equation}
Hence by calculating the Green's functions and using Eq. (\ref{i1})
and (\ref{gd}), we obtain the dynamics conductance which now satisfies
the current conservation and gauge invariance conditions. We further 
point out that result (\ref{gab}) is quite general: it is suitable
for AC transport coefficient in general terms of $\omega$. It can also 
be applied to the strongly interacting system such as the Anderson model 
in the large U limit by including the contribution of displacement current.  
For that model the Green's functions $G^<$ and $G^r$ have been obtained by 
Ng in Ref. \onlinecite{ng}.

Result (\ref{gab}) is in agreement with the formula derived using 
SMT, {\it i.e.} Eq. (16) of Ref. \onlinecite{but1}. However 
the quantities involved in Eq. (\ref{gab}) are calculated within
NEGF and we now derive these necessary quantities. In the mean field 
approximation it is 
straightforward to obtain, from Eqs.(\ref{i1}) and the expression
of $I^c_{\alpha}(\omega)$,
\begin{eqnarray}
& &G^c_{\alpha \beta} (\omega) = -q \int \frac{dE}{2\pi} 
Tr\left[ \bar{g}^<_{\beta} ~ (\Sigma^a_{0\alpha}-
\bar{\Sigma}^r_{0\alpha}) +\bar{g}^r_{\beta} ~ \Sigma_{0\alpha}^< 
\right. \nonumber \\
& &\left.- \bar{\Sigma}^<_{0\alpha} ~ \bar{g}^a_{\beta} 
+(\bar{G}^r_0 ~ \bar{\sigma}^<_{\alpha} - 
\bar{\sigma}^<_{\alpha} ~ G^a_0 
+\bar{G}^<_0 ~ \bar{\sigma}^a_{\alpha} 
- \bar{\sigma}^r_{\alpha} ~ G^<_0) \delta_{\alpha \beta} \right]
\label{gc}
\end{eqnarray}
where we have used the abbreviation $g \equiv g(E)$ and 
$\bar{g} \equiv g(E+ \omega)$ etc. In the above equation, $G_0$ and 
$\Sigma^{<,r,a}$ are the equilibrium Green's function and self-energy 
function; $\bar{\sigma}^{<,r,a}$ is the self-energy function due to the 
AC field: $\bar{\sigma}^<_{\alpha} =(iq/\omega)\left[\Gamma_{\alpha} f -
\bar{\Gamma}_{\alpha} \bar{f} ~~ \right]$
and $\bar{\sigma}^{r,a}_{\alpha} = (q/\omega)
\left [\Sigma^{r,a}_{0\alpha} - \bar{\Sigma}^{r,a}_{0\alpha}\right]$,
where $f$ is the equilibrium Fermi distribution function (at time
$t=-\infty$) of the leads far away from the scattering region.
We have assumed that there is no DC bias at time $t=-\infty$.
The lesser, retarded and advanced Green's functions $\bar{g}^{<,r,a}$ 
are defined as 
$\bar{g}^{r,a}_{\alpha} = \bar{G}^{r,a}_0 ~ \bar{\sigma}^{r,a}_{\alpha}
~ G^{r,a}_0$
and
$\bar{g}^<_{\alpha} = \bar{G}^r_0 ~ \bar{\sigma}^<_{\alpha} ~ G^a_0 +
\bar{G}^r_0 ~ \bar{\sigma}^r_{\alpha} ~ G^<_0 + \bar{G}^<_0 ~ 
\bar{\sigma}^a_{\alpha} ~ G^a_0$.
We emphasize that while equilibrium quantities such as the Fermi
function do appear in this formalism, they only represent the ``initial''
condition: the system was in equilibrium at $t=-\infty$. For $t>-\infty$,
the nonequilibrium Green's function approach describes the full
nonequilibrium dynamic process. With the above definitions and quantities, 
the dynamic conductance $G_{\alpha\beta}(\omega )$ can be evaluated 
for a variety of systems near or far from equilibrium. 

\noindent
{\underline{\bf The wideband limit.}}
As a first example, it is useful to derive an analytic expression of 
admittance in the wideband limit. 
Let's consider the case of resonant tunneling through a quantum dot 
with a single energy level $E_0$. In the wideband limit, the steady state 
Green's function $G^r_0$ is
\[G^r_0 = \frac{1}{E-E_0+i\Gamma/2}\ \ .  \]
At zero temperature and through some straightforward algebra 
we obtain from Eq.(\ref{gab}), 
\begin{equation}
G_{\alpha \beta} = 4 ~ [-\frac{\Gamma_{\alpha}\Gamma_{\beta}}{\Gamma^2}
+\frac{\Gamma_{\alpha}}{\Gamma} \delta_{\alpha \beta} ] ~ X
\label{wide}
\end{equation}
where $X$ is the AC conductance of a noninteracting (without Coulomb 
interaction) symmetrical system obtained in Refs. \onlinecite{fu,jacoboni}.
The noninteracting result does not satisfy current conservation and 
gauge invariance, while the formula (\ref{wide}) does: this is due to the 
prefactor of $X$ in Eq. (\ref{wide}). One can confirm, at the linear
order of AC frequency, that the wideband result (\ref{wide}) agrees 
with the scattering matrix theory\cite{but2} if we assume a Breit-Wigner 
form of the scattering matrix.

\noindent
{\underline{\bf Beyond Wideband Limit.}}
As a main result of this work, we now present the dynamic
conductance beyond the wideband limit. For this case
the situation is quite different and we can only obtain results 
numerically. From Eq.(\ref{gab}) and the expression of 
$\bar{\sigma}^<_{\alpha}$ given above, we also note that due to the energy 
dependence of $\Gamma$, in general the dynamic conductance
depends on the Fermi function in more complicated fashion than
simply $df/dE$. In the following let's calculate the admittance 
of a symmetric double-barrier resonant tunneling structure which 
consists of two $\delta$-function barriers with the same 
strength $V_o$. We solve Eq.(\ref{gab}) numerically using a 
method developed by McLennan {\it et. al}\cite{datta1}. 

Fig. (1) plots $G_{\alpha\beta}(E,\omega)$ as a function of the 
electron Fermi energy E for three values of $\omega$. Both 
real and imaginary part of $G_{21}$ show three peaks due
to resonance transmission through the quantum well. For very small
$\omega$, {\it e.g.} $\omega=0.011$ (solid line), the real
part of $G_{21}(E,\omega)$ essentially coincides with the DC
conductance (not shown); and the imaginary part resembles the 
emittance obtained from SMT\cite{yip}. Substantial deviations
is observed from these known limits when $\omega$ is larger.
A finite $\omega$ tends to smear out the resonance behavior of the 
real part of $G_{21}$. In Fig. (2) we plot $G_{21}(E_f,\omega)$
versus $\omega$ for $E_f\approx 0.15$ which is near the first 
resonance energy of Fig. (1). The extra peaks of Fig. (2)
are due to photon assistant tunneling: the electron can absorb
a photon of appropriate energy and exit from the tunnel structure
at another resonance level. This picture is confirmed by the inset
of Fig. (2) where the energy difference and the frequency difference
between resonances is found to be roughly linear functions of each 
other, with some deviation from perfect linear behavior coming from
uncertainties due to the width of the peaks. Such a linear
behavior has been observed in the experiment of Ref. 
\onlinecite{oosterkamp}. Finally, in Fig. (3) we plot 
$G_{21}(E_f,\omega)$ versus $\omega$ for double-barriers with
infinite height: for this case there is no DC current which can
flow through and the system becomes a parallel plate capacitor. 
Due to displacement field, our formalism 
predicts a non-zero $G_{21}(E_f, \omega)$. We note that not only the 
imaginary part of $G_{21}$ is non-zero, even the real part is 
non-zero. The latter effect gives rise to the charge relaxation 
resistance and is due to contacts implicitly assumed in the 
transport formalisms\cite{but5}. The imaginary part of 
$G_{21}$ for this system, dividing $\omega$, is just the 
$\omega$-dependent charge response coefficient. It gives the
electrochemical capacitance of the system when 
$\omega\rightarrow 0$. The oscillatory behavior can be traced 
back to the retardation effect of Maxwell 
equations, namely, the Helmhotz equation describing the 
{\it characteristic potential}\cite{but2,ma1} has a term proportional
to $\omega^2$ which modifies the value and effective sign 
of the induced charge at large $\omega$, leading to an oscillatory
charge response.

In summary, we have provided a solution to the problem of displacement 
current partition for multi-probe mesoscopic conductors within NEGF.
This allows us to obtain 
dynamic conductance in general terms of the AC frequency and the result
satisfies the current conservation and gauge invariance conditions.
While for special cases ({\it e.g.} the wide band limit) analytic
results can be obtained, in general the dynamic conductance 
formula can be evaluated with well established numerical 
methods. 
While the nonequilibrium Green's function formalism can be applied to
situations far from equilibrium, near equilibrium it must reduce to
the same results as obtained from the scattering matrix theory.  For the
external response, namely the conductance due to particle current,
Ref. \onlinecite{datta} has provided this connection. Including the 
displacement current we have shown, using an explicit example of double 
barrier tunneling diode,  that nonequilibrium Green's function formula 
indeed recovers the scattering matrix result in the wideband limit where the
dynamic conductance become a Fermi surface quantity.

After the submission of our paper, we became aware of a recent
paper\cite{pedersen} which treats photon assisted tunneling in the SMT 
formulation. In this paper, the internal potential
has been considered explicitly and the gauge invariance is satisfied. 

\noindent
{\bf Acknowledgments.}
We gratefully acknowledge support by a RGC grant from the SAR Government 
of Hong Kong under grant number HKU 7112/97P, and a CRCG grant from the 
University of Hong Kong. H. G is supported by NSERC of Canada and FCAR 
of Qu\'ebec. We thank the Computing Center of The University of Hong 
Kong for computational facilities.

\section*{Figure Captions}

\begin{itemize}

\item[{Fig. (1)}] The real part of
$G_{21}(E,\omega)$ as a function of $E$ for
three values of $\omega$. Solid line: $\omega=0.01135$;
dotted: $\omega=0.1022$; dashed: $\omega=0.193$.  Inset: the
corresponding imaginary part. The barriers' strength $V_o=8.81$,
and the quantum well width is $100$\AA. Units of $G_{21}$ is 
$2e^2/h$, of energy $E$ is $eV$. We have set $\hbar=1$. 

\item[{Fig. (2)}] $G_{21}(E_f,\omega)$ as a function of $\omega$,
with $E_f = 0.1533$. Solid: read part; dotted: imaginary part.
The peaks are due to photon assisted tunneling.  Inset: 
a roughly linear relationship between the photon energy and the
resonance energy. Other system parameters and units
are the same as those of Fig. (1).

\item[{Fig. (3)}] $G_{21}(E_f,\omega)$ of a parallel plate 
capacitor as a function of $\omega$ with $E_F=0.1533$. 
The solid (dotted) line is the real (imaginary) part. 
Other system parameters and units are the same as those of Fig. (1).
\end{itemize}

\begin{thebibliography}{00}
\bibitem{foot}
The theory can be generalized to the nonlinear DC case. However even in 
the linear voltage case, a finite frequency can still drive the
system our of equilibrium.
\end{thebibliography}
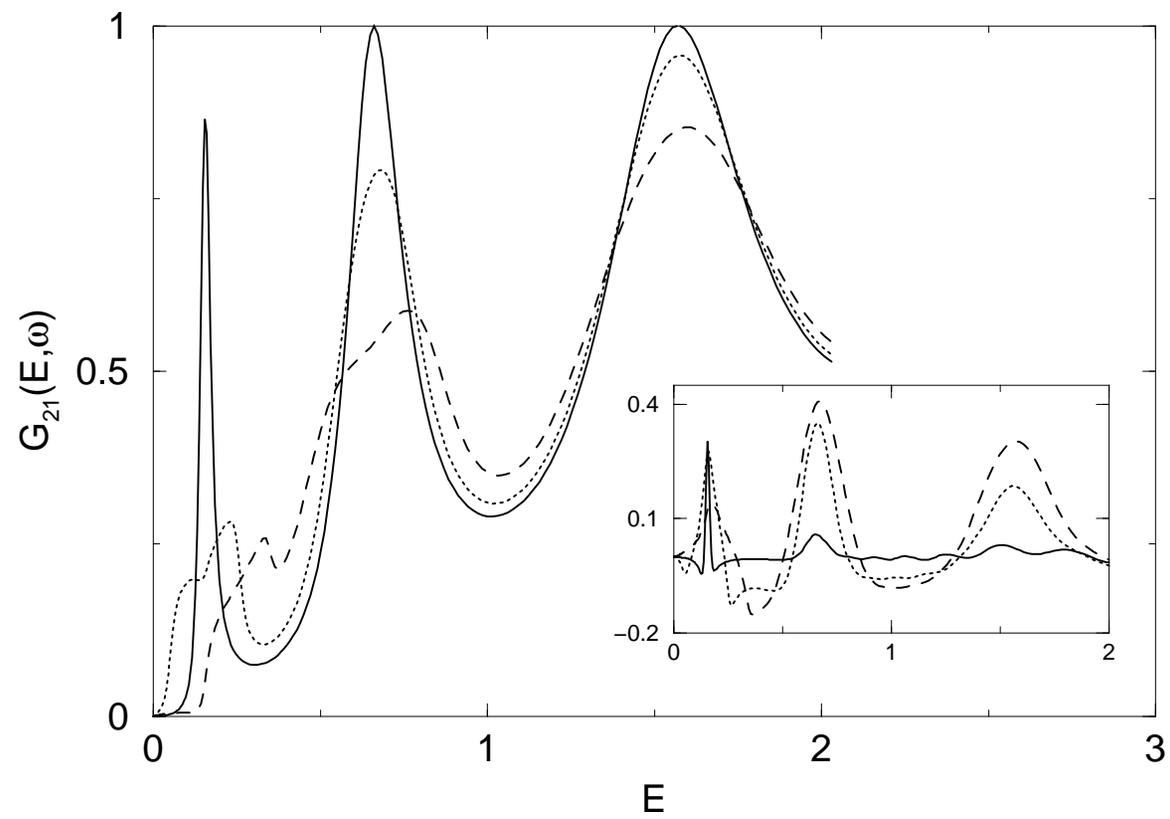
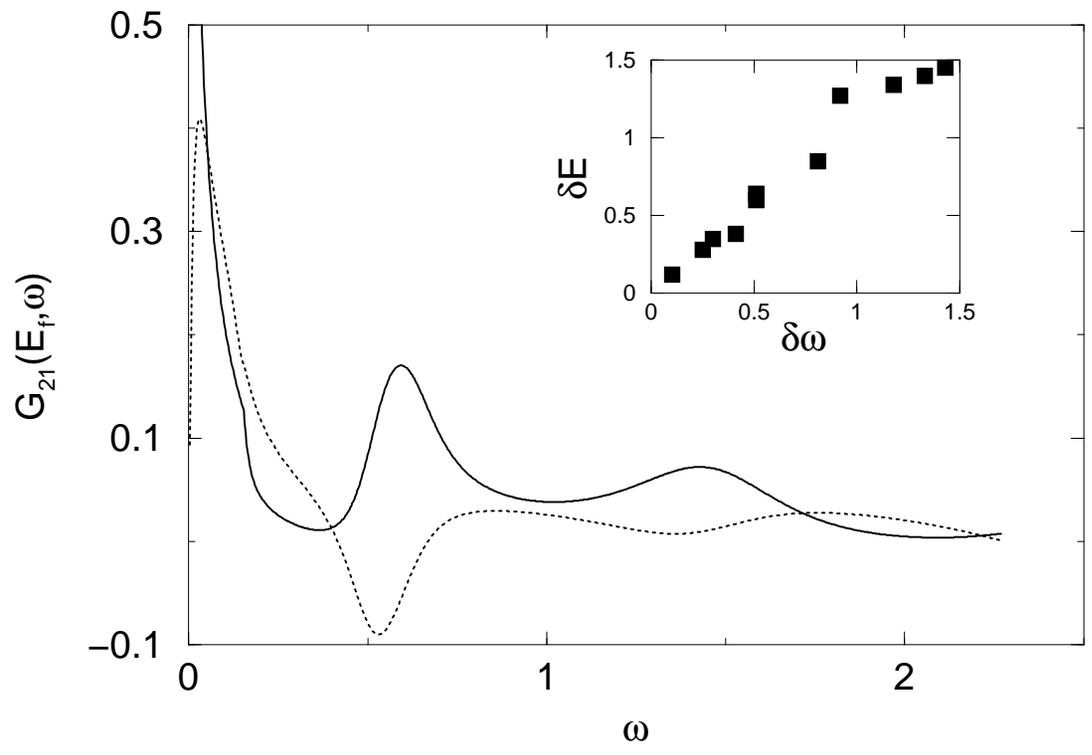
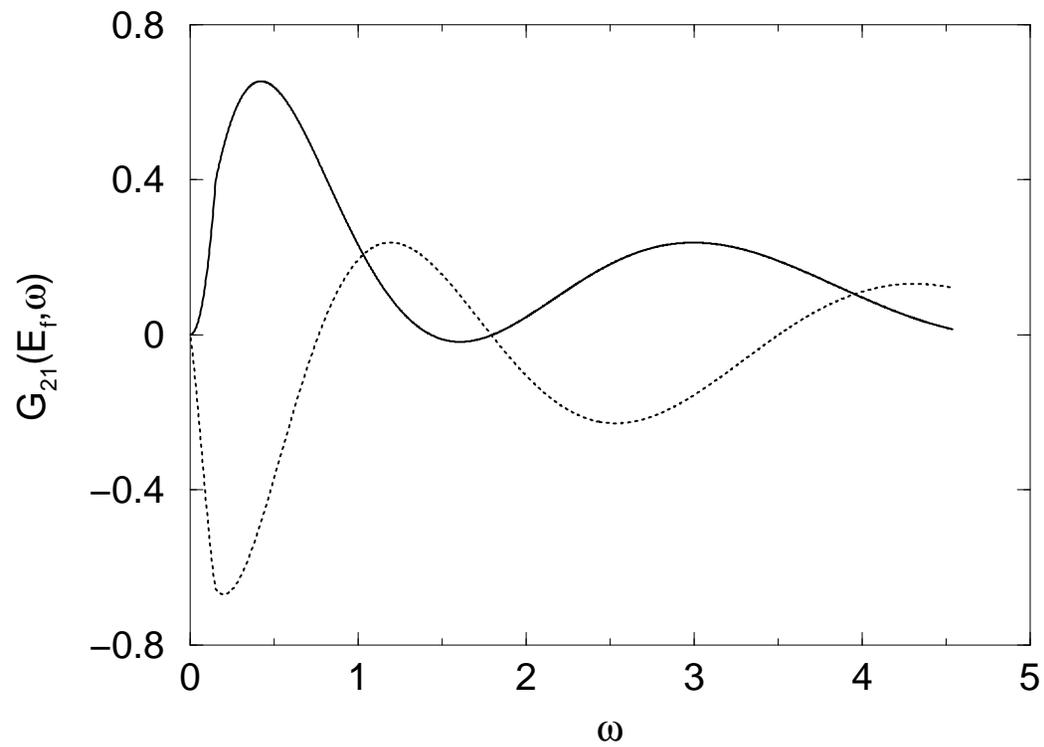
\end{document}